\documentclass[aps,showpacs,preprintnumbers,amsmath,amssymb]{revtex4}
\usepackage{graphics,epsfig}
\usepackage{graphicx}
\usepackage{dcolumn}
\usepackage{bm}

\begin{document}
\baselineskip=0.8 cm
\title{{\bf Influence of Lorentz violation on Dirac quasinormal modes in the Schwarzschild black hole spacetime}}
\author{Songbai Chen}
\email{csb3752@hotmail.com}
\affiliation{Department of Physics,
Fudan University, Shanghai 200433, P. R. China
 \\ Institute of Physics and  Department of Physics,
Hunan Normal University,  Changsha, Hunan 410081, P. R. China }

\author{Bin Wang}
\email{wangb@fudan.edu.cn} \affiliation{Department of Physics, Fudan
University, Shanghai 200433, P. R. China}

\author{ Rukeng Su}
\email{rksu@fudan.ac.cn}
 \affiliation{China Center of Advanced Science and Technology (World Laboratory),
P.B.Box 8730, Beijing 100080, People¡¯s Republic of China
\\ Department of Physics, Fudan University, Shanghai 200433, P. R. China}

\vspace*{0.2cm}
\begin{abstract}
\baselineskip=0.6 cm
\begin{center}
{\bf Abstract}
\end{center}

Using the third-order WKB approximation and monodromy methods, we
investigate the influence of Lorentz violating coefficient $b$
(associated with a special axial-vector $b_{\mu}$ field ) on Dirac
quasinormal modes in the Schwarzschild black hole spacetime. At
fundamental overtone, the real part decreases linearly as the
parameter $b$ increases. But the variation of the imaginary part
with $b$ becomes more complex. For the larger multiple moment $k$,
the magnitude of imaginary part increases with the increase of
$b$, which means that presence of Lorentz violation makes Dirac
field damps more rapidly. At high overtones, it is found that the
real part of high-damped quasinormal frequency does not tend to
zero, which is quite a different from the asymptotic Dirac
quasinormal modes without Lorentz violation.
\end{abstract}

 \maketitle
\newpage
\vspace*{0.2cm}
\section{Introduction}

Since  Lorentz invariance was discovered, it has been great of
importance in many fields of the fundamental physics, such as the
Einstein's special relativity theory, particle physics and high
energy physics. However, the recent development of unified gauge
theories and the observation of high energy cosmic rays
\cite{GTV}\cite{MTA}\cite{SRC} imply that Lorentz symmetry is only
an approximate symmetry of nature and may be spontaneously broken in
the more fundamental physics defined in a higher scale of energy.
Obviously, the existence of Lorentz violation will make a great
influence on the fundamental physics and lead to many subjects need
to be reconsidered. Therefore, a great deal of effort has been
attracted to study Lorentz violation in the different fields
\cite{NAH}-\cite{Ferreira06}.

One of interesting theory models with Lorentz violation is the
Standard Model Extension\cite{AKO1}\cite{AKO2}\cite{AKO3}. It offers
a consistent theoretical framework which includes the standard model
and allows for small violations of Lorentz and CPT symmetry. This
small spontaneous breaking of Lorentz symmetry maybe arise from the
presence of nonzero vacuum expectation values for Lorentz tensors
defined in an underlying theory. A straightforward method of
implementing Lorentz violation in the curve spacetime is to imagine
the existence of a tensor field with a non-vanishing expectation
value and couple this tensor to gravity or matter fields. In
gravitational theories, vierbein formalism is used widely because it
can build a link between the covariant components $T_{\mu\nu\cdots}$
of a tensor field in a coordinate basis and the corresponding
covariant components $T_{ab\cdots}$ of the tensor field in a local
Lorentz frame. The link can be described by $
T_{\mu\nu\cdots}=e^{a}_{\;\; \mu}e^{b}_{\;\; \nu}\cdots
T_{ab\cdots}$, where vierbein $e^a_{\;\;\mu}$ is defined by $
g_{\mu\nu}=\eta_{ab}e^a_{\;\;\mu}e^a_{\;\;\nu}$. Moreover, it is
also found that vierbein formalism can deal with local Lorentz
transformation and diffeomorphisms, which are two basic types of
spacetime transformations in gravitational theories. Bluhm and
Kostelecky\cite{AKO4} make use of this tool and find that any
violation of diffeomorphism invariance via vacuum values of vectors
or tensors breaks local Lorentz invariance. The converse is also
true. For example, if a spacetime vector $b_{\mu}$ acquires a fixed
vacuum expectation value $\langle b_{\mu}\rangle$, which breaks
diffeomorphisms, then the associated local vector $b_{a}$ as given
by contraction with the inverse of vierbein  also acquires a fixed
vacuum expectation value $\langle b_a\rangle$. The presence of
quantity $\langle b_a\rangle$ breaks local Lorentz symmetry.
Therefore, it is natural for us to adopt vierbein formalism to study
of Lorentz violation in the curve spacetime.

Adopting vierbein formalism, the fermion partial action $S_{\psi}$
in Standard Model Extension can be explicitly expressed as
\begin{eqnarray}
S_{\psi}=\int d^4x\sqrt{-g}(\frac{1}{2}i e^{\mu}_a \overline{\psi}
\Gamma^a \overleftrightarrow{D_{\mu}}\psi-\overline{\psi} M^*
\psi),\label{act}
\end{eqnarray}
where $e^{\mu}_{\;\;a}$ is the inverse of the vierbein
$e^a_{\;\;\mu}$. The symbols $\Gamma^a$ and $M^*$ are
\begin{eqnarray}
\Gamma^a\equiv \gamma^a-c_{\mu\nu}e^{\nu a}e^{\mu}_{\;\;
b}\gamma^b-d_{\mu\nu}e^{\nu a}e^{\mu}_{\;\;
b}\gamma_5\gamma^b-e_{\mu}e^{\mu a}-i f_{\mu}e^{\mu
a}\gamma_5-\frac{1}{2}g_{\lambda\mu\nu}e^{\nu a}e^{\lambda}_{\;\;
b}e^{\mu}_{\;\; c}\sigma^{bc}\label{gam1},
\end{eqnarray}
and
\begin{eqnarray}
M^*\equiv m +i m_5\gamma_5+a_{\mu}e^{\mu}_{\;\;a}\gamma^a
+b_{\mu}e^{\mu}_{\;\;a}\gamma_5\gamma^a+\frac{1}{2}H_{\mu\nu}e^{\mu}_{\;\;a}e^{\nu}_{\;\;b}\sigma^{ab}.\label{gam2}
\end{eqnarray}
The first terms of Eqs.(\ref{gam1}) and (\ref{gam2}) lead to the
usual Lorentz invariant kinetic term and mass for the Dirac field.
The parameters $a_{\mu}$, $b_{\mu}$, $c_{\mu\nu}$, $d_{\mu\nu}$,
$e_{\mu}$, $f_{\mu}$, $ g_{\lambda\mu\nu}$ , $H_{\mu\nu}$ are
Lorentz violating coefficients which arise from nonzero vacuum
expectation values of tensor quantities and comprehensive describe
effects of Lorentz violation on  the behavior of particles coupling
to these tensor fields. All of coefficients can be constrained as
the real numbers if the action (\ref{act}) is hermitian. In
generally, they are functions of position. The two terms involving
the couplings $a_{\mu}$ and $b_{\mu}$ are CPT odd, which have been
extensively studied in connection with Lorentz- and CPT-violating
probing experiments including comparative studies of cyclotron
frequencies of trapped-atoms \cite{Bluhm97}, clock comparison tests
\cite{Bluhm02}, spectroscopic comparison of hydrogen and
antihydrogen \cite{Kostelecky99}, analysis of muon anomalous
magnetic moment\cite{Bluhm00}, study of macroscopic samples of
spin-polarized solids \cite{Bluhm001}, and so on.

On the other hand, black hole is another interesting object in the
modern fundamental physics. It is believed widely that the study of
black hole may lead to a deeper understanding of the relationship
among the general relative theory, quantum mechanics, thermodynamics
and statistics. This means that black hole physics play an important
role in the fundamental physics. However, at present whether black
holes exist in our universe or not is still unclear. A recent
investigation shows that quasinormal modes can provide a direct way
to identify black hole existence in our universe because that they
carry the characteristic information of black holes
\cite{Chandrasekhar}\cite{Regge}. Moreover, it is also found that
quasinormal modes have a close connection with the AdS/CFT
correspondence \cite{Witten}\cite{Maldacena}\cite{Kalyana} and the
loop quantum gravity \cite{Hod98}\cite{Dreyer03}. Thus, the study of
quasinormal modes in black hole spacetimes has become appealing in
recent years \cite{Cardoso}-\cite{Jing2}.

Since both Lorentz violation and the quasinormal modes are hot
topics in physics at present, it is natural to raise a question
whether Lorentz violation affects the quasinormal modes of black
holes. From the action (\ref{act}), we can obtain that due to
presence of Lorentz violating coefficients Dirac equation must be
modified and then its quasinormal frequencies in the black hole
spacetimes should be changed. However, to my best knowledge it is
still an open question how Lorentz violation affects the properties
of quasinormal modes in the background of black hole spacetimes.
Obviously, different Lorentz violating coefficients (i.e. different
types of breaking of Lorentz symmetries) have different effects on
quasinormal modes. In this paper, we just consider Dirac equation
with a modified term containing Lorentz violating coefficient $b$
which arises from the nonzero vacuum expectation value of a special
axial vector field $b_{\mu}$ \cite{AKO2}\cite{AKO3} and calculate
quasinormal modes of massless Dirac fields in the Schwarzschild
black hole spacetime. Our result shows that in this case both
fundamental and high overtones quasinormal frequencies depend on the
Lorentz violating coefficient $b$.

The organization of this paper is as follows. In Sec.II, we derive
the equation of massless Dirac field coupled with the special axial
vector field $b_{\mu}$ in the Schwarzschild black hole spacetime. In
Sec.III, we evaluate the fundamental overtones quasinormal
frequencies of the Dirac perturbational field by using the
third-order WKB approximation\cite{BC}\cite{SC}\cite{SI}. In Sec.IV,
we adopt the monodromy technique \cite{Motl03}\cite{Andersson} and
study the high-damped quasinormal frequencies. The last section is
devoted to a summary.

\vspace*{0.2cm}
\section{The Dirac equation with  Lorentz violation associated with an axial-vector $b_{\mu}$ field }

According to the variation of $\overline{\psi}$ in the action
(\ref{act}), we find that the massless Dirac equation only
containing the CPT and Lorentz covariance breaking kinetic term
associated with an axial-vector $b_{\mu}$ field in the curve
spacetime can be expressed as
\begin{eqnarray}
[i\gamma^ae_{a}^{\;\;\mu}(\partial_{\mu}+\Gamma_{\mu})-b_{\mu}e_{a}^{\;\;\mu}\gamma_{5}\gamma^a]\Psi=0,\label{deq1}
\end{eqnarray}
where
\begin{eqnarray}
\gamma^{0}=\left(
\begin{array}{c}I\;\;\;\;\;\;\;\;\;\;0\\
\;0 \;\;\;\;\;\; -I
\end{array}\right),\;\;\;\;\;
\gamma^{i}=\left(
\begin{array}{c}\;\;0\;\;\;\;\;\;\;\;\;\;\sigma^{i}\\
-\sigma^{i} \;\;\;\;\;\; 0
\end{array}\right),\;\;\;\;\;
\gamma_{5}=i\gamma^{0}\gamma^{1}\gamma^{2}\gamma^{3}=\left(
\begin{array}{c}\;\;0\;\;\;\;\;\;I\\
I \;\;\;\;\;\; 0
\end{array}\right).
\end{eqnarray}
Since the Lorentz violation is very small, it is reasonable for us
to assume that the axial-vector $b_{\mu}$ field does not change the
background metric. For convenience, we take $b_{\mu}$ as a non-zero
timelike vector $(\frac{b}{r^2},0,0,0)$, where $b$ is a constant. In
the Schwarzschild spacetime, the vierbein can be defined as
\begin{eqnarray}
e^a_{\;\;\mu}=(\sqrt{1-\frac{2M}{r}},
\frac{1}{\sqrt{1-\frac{2M}{r}}}, \;\; r, \;\;
r\sin{\theta})\label{vier1}.
\end{eqnarray}
Setting
$\Psi=(1-\frac{2M}{r})^{-\frac{1}{4}}(\sin{\theta})^{-\frac{1}{2}}
\Phi $ and substituting Eq.(\ref{vier1}) into Eq.(\ref{deq1}), the
Dirac equation (\ref{deq1}) can be simplified as
\begin{eqnarray}
\left[\frac{\gamma^0}{\sqrt{1-\frac{2M}{r}}}\frac{\partial}{\partial
t}+\sqrt{1-\frac{2M}{r}}\gamma^{1}(\frac{\partial }{\partial
r}+\frac{1}{r})+\frac{\gamma^2}{r}\frac{\partial }{\partial
\theta}+\frac{\gamma^3}{r\sin{\theta}}\frac{\partial}{\partial
\varphi}+\frac{b}{r^2\sqrt{1-\frac{2M}{r}}}\gamma_{5}\gamma^{0}\right]\Phi=0.\label{deq3}
\label{deq2}
\end{eqnarray}
If we define a tortoise coordinate
\begin{eqnarray}
r_{*}=r+2M\ln{(\frac{r}{2M}-1)},
\end{eqnarray}
and the ansatz
\begin{eqnarray}
\Phi=\left(\begin{array}{c}\frac{iG^{(\pm)}(r)}{r}\phi^{\pm}_{jm}(\theta,\varphi)\\
\\ \frac{F^{(\pm)}(r)}{r}\phi^{\mp}_{jm}(\theta,\varphi)
\end{array}\right)e^{-i\omega t}
\end{eqnarray}
with
\begin{eqnarray}
\phi^{+}_{jm}&=&\left(\begin{array}{c}
\sqrt{\frac{j+m}{2j}}Y^{m-1/2}_{l} \\
\\ \sqrt{\frac{j-m}{2j}}Y^{m+1/2}_{l}
\end{array}\right),\;\;\;\;\;\; \;\;\;\;\;\;\;\;\;\;\;
k=j+\frac{1}{2},\;\;\; j=l+\frac{1}{2},\nonumber\\ \\
\phi^{-}_{jm}&=&\left(\begin{array}{c}
\sqrt{\frac{j-m+1}{2j+2}}Y^{m-1/2}_{l}\\
\\ -\sqrt{\frac{j+m+1}{2j+2}}Y^{m+1/2}_{l}
\end{array}\right),\;\;\;\;\;\;\;\;\; \;\;k=-(j+\frac{1}{2}),\;\;\;
j=l-\frac{1}{2},\nonumber
\end{eqnarray}
we find that the case for (+) and $(-)$ in the functions $F^{\pm}$
and $G^{\pm}$ can be put together and Eq.(\ref{deq3}) can be
rewritten as
\begin{eqnarray}
\frac{d}{dr_*}\left(\begin{array}{c}G\\
F\end{array}\right)+\left(\frac{k}{r}\sqrt{1-\frac{2M}{r}}-\frac{b}{r^2}\right)\left(\begin{array}{c}I\;\;\;\;\; 0\\
0\;\;-I\end{array}\right)\left(\begin{array}{c}G\\
F\end{array}\right)=\left(\begin{array}{c}0\;\;\;\;\ \omega\\
-\omega \;\;\;\;\;\;0\end{array}\right)\left(\begin{array}{c}G\\
F\end{array}\right).\label{deq4}
\end{eqnarray}
It is very easy to find the decoupled equations for variables $F$
and $G$ can be expressed as
\begin{eqnarray}
&& \frac{d^2 F}{dr^2_{*}}+(\omega^2-V_1)F=0, \label{eqv1}\\ &&
\frac{d^2G}{dr^2_{*}}+(\omega^2-V_2)G=0,
\end{eqnarray}
with
\begin{eqnarray}
V_{1,2}&=&\pm\frac{d W}{dr_*}+W^2\nonumber\\ &=& \frac{|k|
\Delta^{\frac{1}{2}}}{r^4}[|k|\Delta^{\frac{1}{2}}\mp(r-3M)-2b]+\frac{b^2\pm2b(r-2M)}{r^4},\label{pot1}
\end{eqnarray}
where $W=\frac{k}{r}\sqrt{1-\frac{2M}{r}}-\frac{b}{r^2}$ and
$\Delta=r(r-2M)$. It is obvious that the potentials $V_{1}$ and
$V_{2}$ are related to the coefficient $b$ of Lorentz violations,
which means that the Dirac quasinormal modes should depend on
Lorentz violation. If we set $b=0$, Eqs.(\ref{eqv1})-(\ref{pot1})
give the results of general Dirac fields in the Schwarzschild black
hole spacetime\cite{Cho}. Moreover, it is well known that the
potentials $V_{1}$ and $V_{2}$ possess the same spectra of
quasinormal frequencies because that they are supersymmetric
partners derived from the same superpotential $W$. In the following,
we therefore just make use of Eq.(\ref{eqv1}) to evaluate the Dirac
quasinormal frequencies and write $V_1$ as $V$.

\section{the fundamental Dirac quasinormal modes with Lorentz violating coefficient $b_{\mu}$ }

In order to study the relationship between quasinormal frequencies
and the coefficient $b$, we can take $M=1$.
\begin{figure}[h]
\begin{center}
\includegraphics[width=5cm]{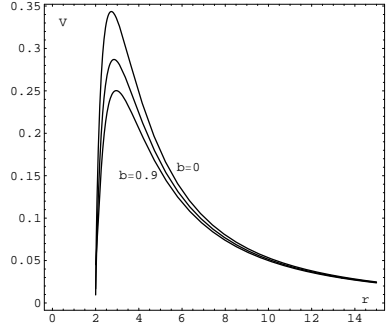}
\caption{Variation of the effective potential for the massless Dirac
field ($k=3$) with the coefficient $b$ of Lorentz violations}
\end{center}
\label{fig1}
\end{figure}
Fig. 1 shows the variation of the effective potential with the
coefficient $b$ of Lorentz violations for fixed $k=3$. From this
figure we can find that as $b$ increases, the peak value of the
potential barrier gets lower and the location of the peak ($r=r_p
\;$) moves along the right.

Let us now evaluate the fundamental quasinormal frequencies for
the massless Dirac field with Lorentz violations by using the
third-order WKB potential approximation, a numerical method
devised by Schutz, Will and Iyer \cite{BC}\cite{SC}\cite{SI}. Due
to its considerable accuracy for lower-lying modes, this method
has been used extensively in evaluating quasinormal frequencies of
various black holes. In this approximate method, the formula for
the complex quasinormal frequencies $\omega$ is
\begin{eqnarray}
\omega^2=[V_0+(-2V^{''}_0)^{1/2}\Lambda]-i(n+\frac{1}{2})(-2V^{''}_0)^{1/2}(1+\Omega),
\end{eqnarray}
where
\begin{eqnarray}
\Lambda&=&\frac{1}{(-2V^{''}_0)^{1/2}}\left\{\frac{1}{8}\left(\frac{V^{(4)}_0}{V^{''}_0}\right)
\left(\frac{1}{4}+\alpha^2\right)-\frac{1}{288}\left(\frac{V^{'''}_0}{V^{''}_0}\right)^2
(7+60\alpha^2)\right\},\nonumber\\
\Omega&=&\frac{1}{(-2V^{''}_0)}\bigg\{\frac{5}{6912}
\left(\frac{V^{'''}_0}{V^{''}_0}\right)^4
(77+188\alpha^2)\nonumber\\&-&
\frac{1}{384}\left(\frac{V^{'''^2}_0V^{(4)}_0}{V^{''^3}_0}\right)
(51+100\alpha^2)
+\frac{1}{2304}\left(\frac{V^{(4)}_0}{V^{''}_0}\right)^2(67+68\alpha^2)
\nonumber\\&+&\frac{1}{288}
\left(\frac{V^{'''}_0V^{(5)}_0}{V^{''^2}_0}\right)(19+28\alpha^2)-\frac{1}{288}
\left(\frac{V^{(6)}_0}{V^{''}_0}\right)(5+4\alpha^2)\bigg\},
\end{eqnarray}
and
\begin{eqnarray}
\alpha=n+\frac{1}{2},\;\;\;\;\;
V^{(s)}_0=\frac{d^sV}{dr^s_*}\bigg|_{\;r_*=r_*(r_{p})} \nonumber,
\end{eqnarray}
$n$ is overtone number.

Substituting the effective potential $V$ into the formula above,
we can obtain the quasinormal frequencies for the Dirac fields
with Lorentz violations.
\begin{table}[h]
\begin{center}
\begin{tabular}[b]{c|c|c|c|c|c}
 \hline \hline
 $b$  & $\omega\ \ \ (k=1)$ &  $\omega \ \ \ (k=2)$
 & $\omega \ \ \ (k=3)$ & $\omega \ \ \ (k=4)$ & $\omega \ \ \ (k=5)$\\ \hline

 0& \;0.176452-0.100109i\; & \;0.378627-0.096542i\; & \;0.573685-0.096320i\; & \;0.767194-0.096276i&\;0.960215-0.096256i
 \\
 0.1& 0.166462-0.100956i&0.367984-0.096909i&0.562847-0.096596i&0.756264-0.096489i&0.949237-0.096431i
 \\
0.2&0.157373-0.101275i&0.357770-0.097175i&0.552294-0.096829i&0.745548-0.096681i&0.938430-0.096592i
 \\
0.3&0.149036-0.101037i&0.347972-0.097342i&0.542020-0.097021i&0.735043-0.096851i&0.927792-0.096739i
\\
0.4&0.141230-0.100325i&0.338572-0.097408i&0.532020-0.097174i&0.724745-0.097000i&0.917321-0.096873i
\\
0.5&0.133666-0.099621i&0.329548-0.097380i&0.522288-0.097288i&0.714652-0.097128i&0.907015-0.096994i
\\
0.6&0.126319-0.101000i&0.320878-0.097266i&0.512819-0.097363i&0.704761-0.097234i&0.896874-0.097101i
\\
0.7&0.122890-0.114215i&0.312532-0.097083i&0.503605-0.097402i&0.695069-0.097319i&0.886894-0.097194i
\\
0.8&-\;-\;-\;-&0.304478-0.096864i&0.494639-0.097404i&0.685573-0.097384i&0.877074-0.097275i
\\
0.9&-\;-\;-\;-&0.296676-0.096668i&0.485914-0.097372i&0.676269-0.097428i&0.867413-0.097342i
\\
1.0&-\;-\;-\;-&0.289080-0.096612i&0.477423-0.097310i&0.667154-0.097453i&0.857907-0.097397i
\\
\hline\hline
\end{tabular}
\end{center}
\caption{The fundamental overtones ($n=0$) quasinormal frequencies
of Dirac field  with Lorentz violating kinetic term associated with
an axial-vector $b_{\mu}$ in the Schwarzschild black hole spacetime
for $k=1 \sim 5$. }
 \end{table}
The fundament modes frequencies for $k=1 \sim 5$ are list in the
Table 1. From Fig.2, we find that for fixed $k$ the real part
almost decrease linearly with the increase with $b$. Moreover,
Fig. 3 tells us that the relationship between the magnitude of the
imaginary parts and $b$ is more complex. For larger $k$, we find
that they increase as $b$ increases, which means that presence of
Lorentz violation makes Dirac oscillation damps more rapidly.
\begin{figure}[ht]
\begin{center}
\includegraphics[width=5cm]{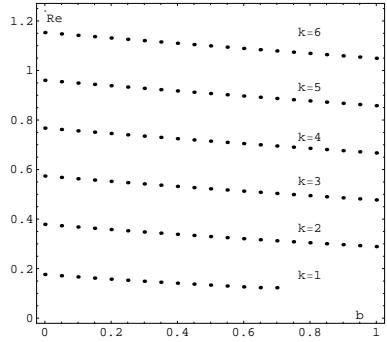}
\caption{Variation of the real parts of Dirac quasinormal
frequencies with the coefficient $b$ of Lorentz violations. }
 \end{center}
 \label{fig2}
 \end{figure}
\begin{figure}[ht]
\begin{center}
\includegraphics[width=5cm]{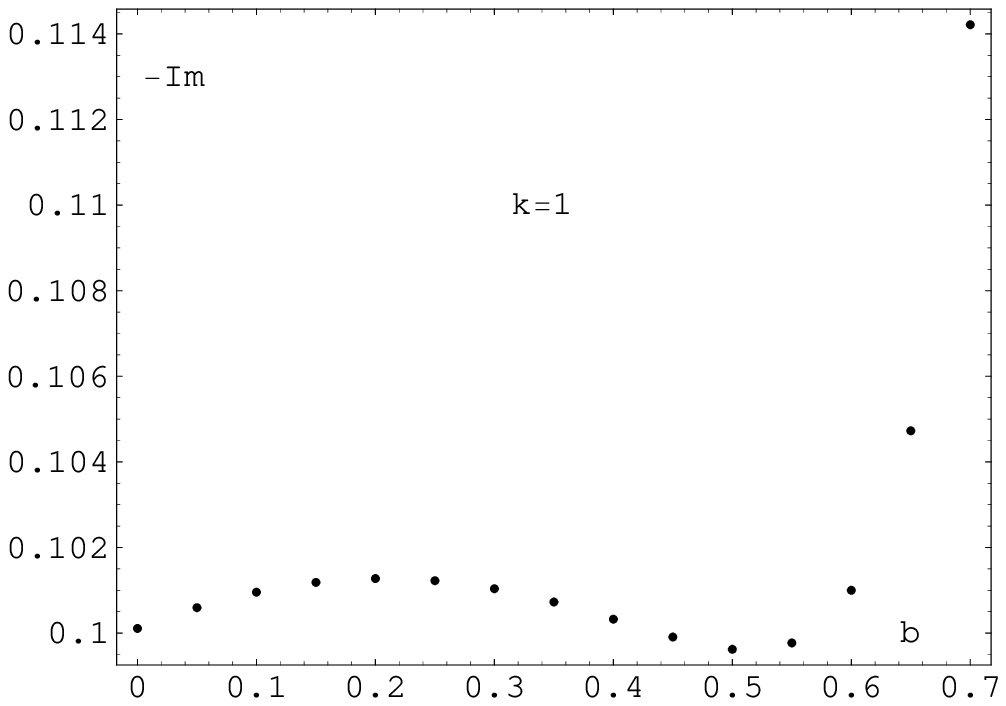}\ \ \ \ \
\includegraphics[width=5cm]{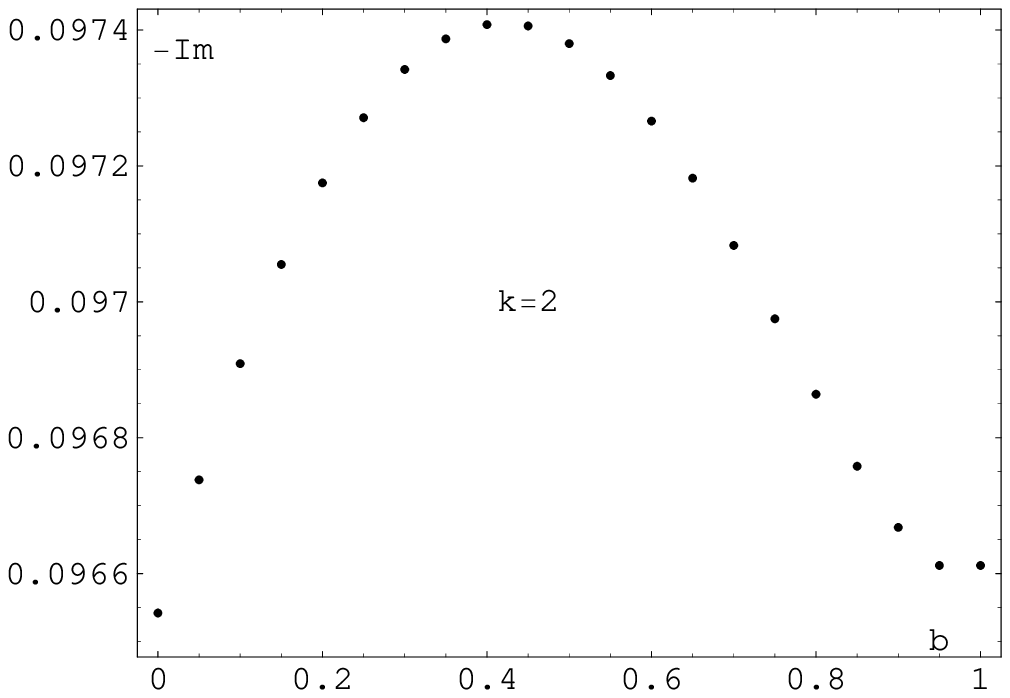}\\
\includegraphics[width=5cm]{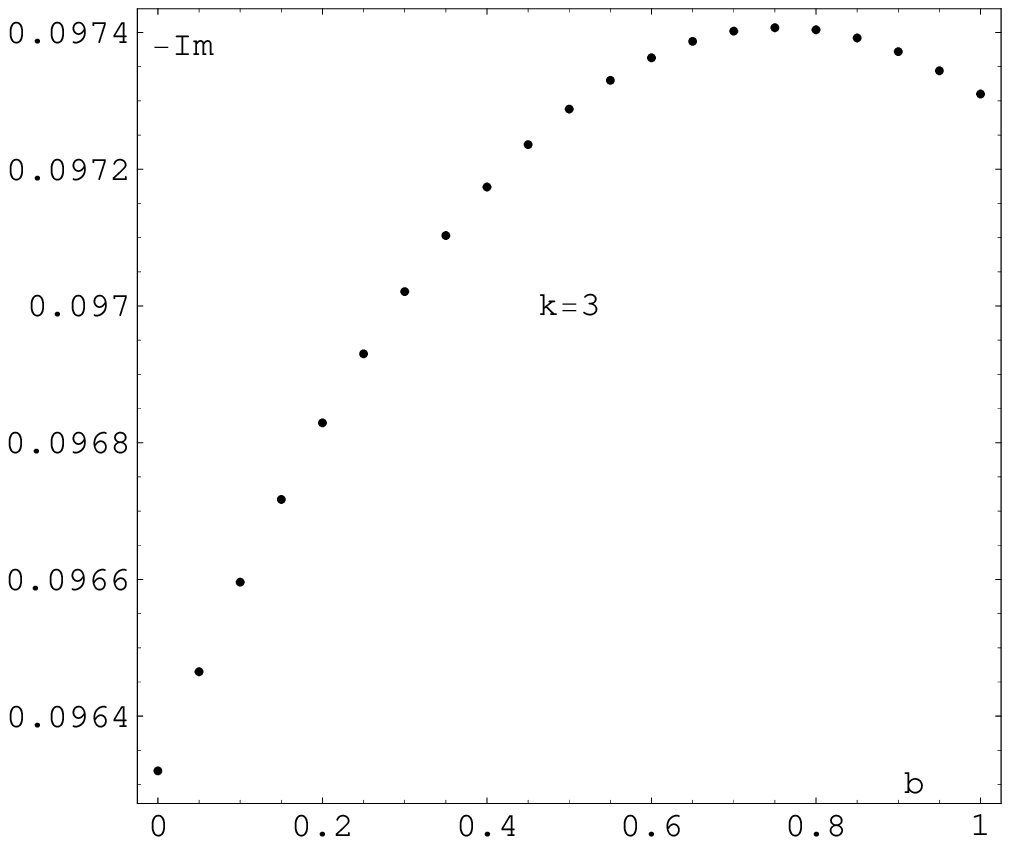}\ \ \ \ \ \includegraphics[width=5cm]{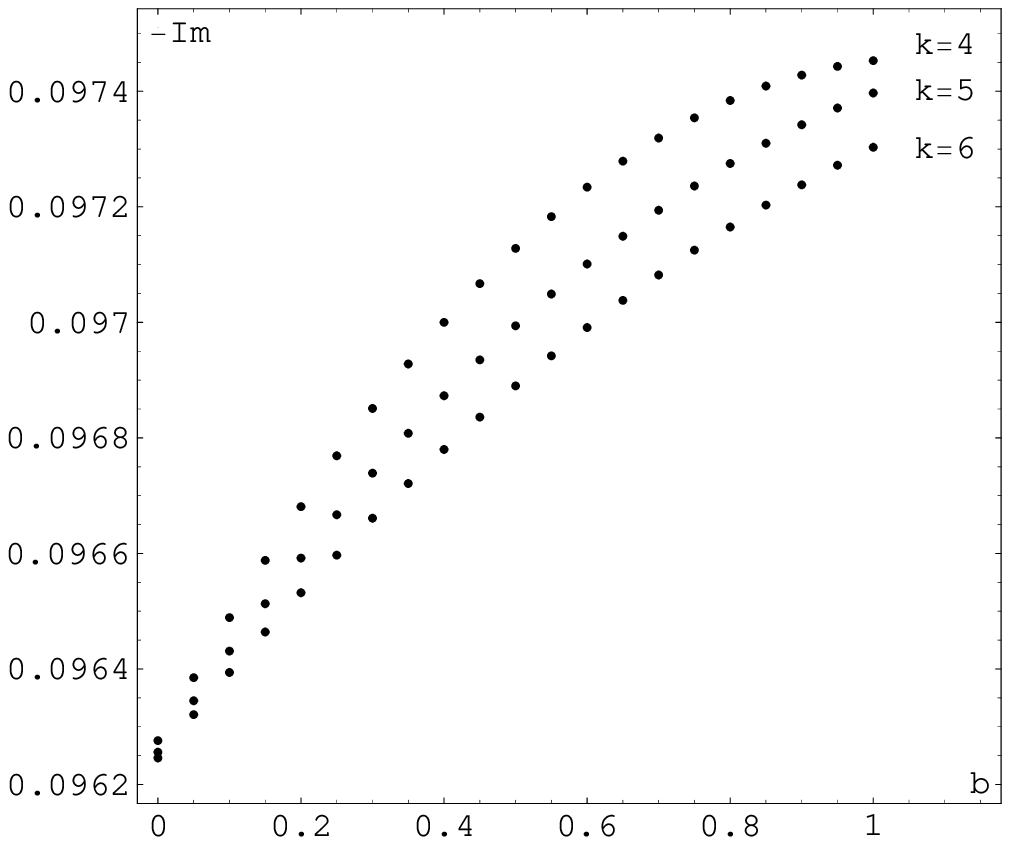}
\caption{Variation of the imaginary parts of Dirac quasinormal
frequencies with the coefficient $b$ of Lorentz violations.}
\end{center}
\label{fig3}
\end{figure}

\section{the high-damped Dirac quasinormal modes with Lorentz violating coefficient $b_{\mu}$}

Motivated by Hod conjecture\cite{Hod98},  a great deal of effort has
been devoted to the study of the high-damped quasinormal modes in
the different black hole spacetimes because that Hod's conjecture
suggests that there maybe exist a connection between the high-damped
quasinormal frequencies and quantum gravity. In this section, we
adopt to the monodromy method\cite{Motl03}\cite{Andersson} and
investigate the high-damped Dirac quasinormal modes with Lorentz
violations in the Schwarzschild black hole spacetime. Our purpose is
to probe whether Lorentz violations affects Hod conjecture. As in
Ref.\cite{Motl03}, after selecting the contour $L$ as shown in
Fig.4, we can calculate the global monodromy around the contour $L$.
In the neighborhood of the event horizon $r=2M$, the effective
potential $V$  and the solution of Eq.(\ref{eqv1}) can be
approximated as
\begin{eqnarray}
V\sim \frac{b^2}{(2M)^4},\;\;\;\;\;\;\;\;\;\;\;\;\;\;
 \phi(r)\sim e^{i \sqrt{\omega^2-\frac{b^2}{(2M)^4}}\;z}.
\end{eqnarray}
Since the only singularity of $\phi(r)$ or $e^{-i \omega z}$
inside the contour occurs at the point $r=2M$. After a full
clockwise round trip, $\phi(r)$ acquires a phase
$e^{\frac{\pi\sqrt{\omega^2-\frac{b^2}{(2M)^4}}}{\kappa}}$, while
$e^{-i\omega z}$ acquires a phase
    $e^{-\frac{\pi\omega}{\kappa}}$. So the coefficient of
  $e^{-i\omega z}$ in the asymptotic of $\phi(r)$ must be multiplied by
$e^{\frac{\pi(\omega+\sqrt{\omega^2-\frac{b^2}{(2M)^4}})}{\kappa}}$.
\begin{figure}[ht]
\begin{center}
\includegraphics[width=5cm]{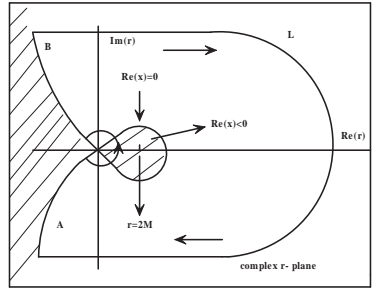}
\caption{The complex $r$-plane and the contour $L$. The regions
with the hachures denote the area $Re(x)<0$.}
\end{center}
\label{fig4}
\end{figure}

Moreover, we find that the behaviors of the tortoise coordinate
$r_*$ and of the effective potential $V$ in Eq.(\ref{eqv1}) near
the singular point $r=0$ are
\begin{eqnarray}
r_*&\sim &-\frac{r^2}{4M},\nonumber\\
 V&\sim&\frac{b^2-4Mb}{r^{4}}=-\frac{1-j^2}{4r_*^{2}},
\end{eqnarray}
where $j=1-\frac{b}{2M}$. As in Ref.\cite{Motl03}, according to
the local monodromy around the singular point $r=2M$, we find that
the monodromy around the contour $L$ must multiply the coefficient
of $e^{-i\omega z}$ by a factor $-(1+2\cos{\pi j})$.

Comparing the local and global monodromy,  we can obtain directly
the high-damped quasinormal frequencies formula for Dirac fields
with Lorentz violations in the Schwarzschild black hole spacetime
\begin{eqnarray}
e^{\frac{\pi\bigg[\omega+\sqrt{\omega^2-\frac{b^2}{(2M)^4}}\bigg]}{\kappa}}=-[1+2\cos{\pi(
1-\frac{b}{2M})}],\;\;\;\;\; \;\;\;\;\; n\rightarrow
\infty.\label{22}
\end{eqnarray}
Where $\kappa$ is the surface gravity constant of the event horizon
of the black hole. Comparing with mass $M$ of black hole, parameter
$b$ is very small, thus the term $\frac{b^2}{(2M)^4}$ in
Eq.(\ref{22}) can be neglected. The frequency formula for
high-damped quasinormal modes can be further simplified as
\begin{eqnarray}
\omega=T_H\ln{|[1+2\cos{\pi(1-\frac{b}{2M})}]|}-i 2n\pi
T_H,\;\;\;\;\; \;\;\;\;\; n\rightarrow \infty, \label{w12}
\end{eqnarray}
where $T_H$ is the Hawking temperature of Schwarzschild black hole.
It is obvious that Lorentz violating coefficient $b$ affects the
high-damped Dirac quasinormal frequencies.

In the higher dimensional ($D>4$, $D$ is the dimensions of spacetime
) Schwarzschild black hole spacetime,  the axial vector field can be
 taken as $D$-component form $(\frac{b}{r^{D-2}},\underbrace{0, \cdots, \cdots,
 0}_{D-1})$. Similarly, we find that in this case the behaviors of the tortoise coordinate $r_*$ and of
the effective potential $V$ near the singular point $r=0$ are
\begin{eqnarray}
r_*&\sim &-\frac{r^{D-2}}{2(D-2)M},\nonumber\\
 V&\sim&\frac{b^2-2(D-2)M b}{r^{2(D-2)}}=-\frac{1-j^2}{4r_*^{2}},
\end{eqnarray}
where $j=1-\frac{b}{(D-2)M}$. Repeating above operations, we can
obtain that the high-damped Dirac quasinormal frequencies in the $D$
dimensional Schwarzschild black hole spacetime satisfy
\begin{eqnarray}
\omega=T_H\ln{|1+2\cos{\pi[1-\frac{b}{(D-2)M}]}|}-i2n\pi
T_H,\;\;\;\;\; \;\;\;\;\; n\rightarrow \infty. \label{w14}
\end{eqnarray}
It is shown that the asymptotic frequency formula of Dirac
quasinormal modes can be also affected by the coefficient $b$ of
Lorentz violation in the $D$ dimensional Schwarzschild black hole
spacetime. When $b\rightarrow 0$, we find that the real parts of
high-damped Dirac quasinormal frequencies in both cases become zero,
which agrees with the result of Dirac field without Lorentz
violations\cite{Cho2}\cite{Jing2}. According to Hod's idea
\cite{SHUK}, one can obtain that classical ringing frequencies with
an asymptotically vanishing real part correspond to virtual quanta
and the corresponding Dirac transitions in Lorentz invariance are
quantum mechanically forbidden. However, from the formulas
(\ref{w12}) and (\ref{w14}), we find that Dirac quantum transition
is allowable in the Lorentz violation Frame. It implies that the
emergence of Lorentz violation may change the quantum property of
Dirac field.

\section{summary}

Adopting the third-order WKB approximation  and monodromy methods,
we investigated the quasinormal modes of Dirac fields with Lorentz
violating term  associated with a special axial-vector $b_{\mu}$ in
the Schwarzschild black hole spacetime. We find that the coefficient
$b$ of Lorentz violation affects Dirac quasionrmal frequencies. At
fundamental overtone, the real part decreases linearly as the
parameter $b$ increases. But the variation of the imaginary part
with $b$ becomes more complex. For the larger multiple moment $k$,
the magnitude of imaginary part increases with the increase of $b$,
which means that presence of Lorentz violation makes Dirac field
damps more rapidly. Since the imaginary part of quasinormal
frequencies for large multiple number $k$ can be well approximated
as \cite{VFB} $\omega_{I} \sim
\frac{-i}{3\sqrt{3}GM}(n+\frac{1}{2})$, the possible reason for our
result is that the presence of the axial timelike vector field
$b_{\mu}$ with nonzero vacuum expectation values may lead to the
decrease of Newton's gravity constant $G$, which is possible in the
Lorentz violating theories. Moreover, Dirac also supports that
Newton's gravity constant $G$
decrease as increase of the age of universe. At high overtones, the
real parts depend on Lorentz violating coefficient $b$. As
$b\rightarrow 0$, we find the real part of high-damped Dirac
quasinormal frequencies becomes zero, which consists with the result
of Dirac field without Lorentz violations. Moreover, our result also
shows that the emergence of Lorentz violation may change the quantum
property of Dirac field. The effects of other Lorentz violating
coefficients in the action (\ref{act}) on quasinormal modes of black
holes need to be investigated in the future.

\begin{acknowledgments}
We thank the referee for his/her quite useful and helpful comments
and suggestions, which help deepen our understanding of Lorentz
violation and quasinormal modes. This work was partially supported
by NNSF of China, Ministry of Education of China and Shanghai
Education Commission. R. K. Su's work was partially supported by
the National Basic Research Project of China. S. B. Chen's work
was partially supported  by the Hunan Provincial Natural Science
Foundation of China under Grant No.05JJ40012 and Scientific
Research Fund of Hunan Normal University under Grant No.22040639.
\end{acknowledgments}


\newpage
\vspace*{0.2cm}
 \begin{thebibliography}{99}

\bibitem{GTV} Zatsepin G T and Kuzmin V A 1966 \textit{JETP Lett.} {\bf 4} 78

\bibitem{MTA} Takeda M et al 1998 \textit{Phys. Rev. Lett.} {\bf 81} 1163

\bibitem{SRC} Coleman S R and Glashow S L 1999 \textit{Phys. Rev.} D {\bf 59} 116008

\bibitem{NAH} Arkani-Hamed N, Cheng H C, Luty M A and  Muko-hyama S
2004 \textit{JHEP} {\bf 0405} 074

\bibitem{CCJE} Csaki C, Erlich J and Grojean C 2001 \textit{Nucl. Phys.} B {\bf 604} 312

\bibitem{JMC} Cline J M and L. Valcarcel, JHEP {\bf 0403} 032 (2004).

\bibitem{MVL} Libanov M V and Rubakov V A 2005 \textit{Phys. Rev.} D {\bf 72} 123503

\bibitem{AKO1} Kostelecky V A 2004 \textit{Phys. Rev.} D {\bf 69} 105009

\bibitem{AKO2} Colladay D and Kostelecky V A 1997 \textit{Phys. Rev.} D {\bf 55} 6760


\bibitem{AKO3} Colladay D and Kostelecky V A 1998 \textit{Phys. Rev.} D {\bf 58} 116002

\bibitem{AKO4} Bluhm R and Kostelecky V A  2005 \textit{Phys. Rev.} D {\bf 71} 065008


\bibitem{SUK} Kanno S and Soda J 2006 \textit{Phys. Rev.} D{\bf 74} 063505

\bibitem{JAAA}Alfaro J, Andrianov A A, Cambiaso M, Giacconi P and Soldati
R 2006 \textit{Phys. Lett.} B {\bf 639} 586-590

\bibitem{HTM} Belich H, Costa-Soares T, Ferreira Jr M M, Helayel-Neto J A and Mouchereck F M O 2006 \textit{Phys. Rev.}
D{\bf 74} 065009


\bibitem{JHS} Henson J 2006 Macroscopic observables and Lorentz violation in discrete
quantum gravity \textit{Preprint} gr-qc/0604040

\bibitem{MRMT} Martinez M R and Piran T 2006 \textit{JCAP}{\bf 0604}  006

\bibitem{Bluhm97} Bluhm R, Kostelecky V A and Russell N 1997 \textit{Phys. Rev.
Lett.}{\bf 79} 1432

\bibitem{Bluhm02} Bluhm R, Kostelecky V A, Lane C D and Russell N 2002 \textit{Phys.
Rev. Lett.} {\bf 88} 090801

\bibitem{Kostelecky99} Kostelecky V A and Russell N 1999 \textit{Phys. Rev. Lett.}
 {\bf 82} 2254

\bibitem{Bluhm00} Bluhm R, Kostelecky V A and Lane C D 2000 \textit{Phys.
Rev. Lett.} {\bf 84} 1098

\bibitem{Bluhm001} Bluhm R and Kostelecky V A 2000 \textit{Phys. Rev. Lett.}
 {\bf 84} 1381

\bibitem{Ferreira06} Ferreira Jr M M and Mouchereck F M O 2006
Influence of Lorentz- and CPT-violating terms on the Dirac
equation \textit{Preprint} hep-ph/0601018


\bibitem{Chandrasekhar} Chandrasekhar S and Detweiler S L
1975 \textit{Proc. R. Soc. London}  A {\bf 344} 441

\bibitem{Regge} Regge T and Wheeler J A 1957 \textit{Phys. Rev.} {\bf 108}
1063

\bibitem{Witten} Witten E 1998 \textit{Adv. Theor. Math. Phys.}{\bf  2} 253

\bibitem{Maldacena} Maldacena J 1998 \textit{Adv. Theor. Math.
Phys.}{\bf  2} 231

\bibitem{Kalyana} Kalyana S R and Sathiapalan B 1999 \textit{Mod Phys. Lett.
 A} {\bf  14} 2635

\bibitem{Hod98} Hod S 1998 \textit{Phy. Rev. Lett.} {\bf 81}  4293

\bibitem{Dreyer03} Dreyer O 2003 \textit{Phy. Rev. Lett.} {\bf 90}  081301

\bibitem{Cardoso} Cardoso V and Lemos J P S 2001 \textit{Phys. Rev.} D {\bf 63}
124015

\bibitem{Konoplya} Konoplya R A 2002 \textit{Phys. Rev.} D {\bf 66} 084007

\bibitem{Starinets} Starinets A O 2002 \textit{Phys. Rev.} D {\bf 66} 124013

\bibitem{Setare} Setare M R 2003 \textit{Class. Quant. Grav.} {\bf 21} 1453
; 2004 \textit{Phys. Rev.} D {\bf 69} 044016

\bibitem{Natario} Natario J and Schiappa R 2004 \textit{Adv. Theor. Math. Phys.} {\bf
8} 1001-1131

\bibitem{Leaver}  Leaver E W  \textit{Proc. R. Soc. Lond.} {\bf A 402} 285 (1985)
; 1986 \textit{Phys. Rev.} D {\bf 34} 384

\bibitem{VFB} Ferrari V and Mashhoon B 1984 \textit{Phys. Rev.} D{\bf 30} 295-304

\bibitem{Cho} Cho H T 2003 \textit{Phys. Rev.} D {\bf 68}  024003

\bibitem{bin1} Shao C G, Wang B, Abdalla E and Su R K 2005 \textit{Phys. Rev.} D {\bf 71 }
044003

\bibitem{bin2} Wang B, Lin C Y and Molina C 2004 \textit{Phys. Rev.} D {\bf 70} 064025

\bibitem{bin3} Du D P, Wang B and Su R K 2004 \textit{Phys. Rev.} D {\bf 70} 064024

\bibitem{Zhidenko} Zhidenko A 2004 \textit{Class. Quant. Grav.} {\bf 21} 273
\bibitem{Jing1} Jing J L 2004 \textit{Phys. Rev.} D {\bf 69} 084009

\bibitem{Cho2} Cho H T 2006 \textit{Phys. Rev.} D {\bf 73} 024019

\bibitem{Jing2} Jing J L 2004 \textit{Phys. Rev.} D {\bf 70}065004;
2005 \textit{Phys. Rev.} D {\bf 71}  124011; 2005 \textit{Phys.
Rev.} D {\bf 71} 124006

\bibitem{BC} Schutz B F and Will C M 1985 \textit{Astrophys. J. Lett. Ed.}
{\bf 291} L33

\bibitem{SC} Iyer S and Will C M 1987
\textit{Phys. Rev.}D {\bf 35} 3621

\bibitem{SI} Iyer S 1987 \textit{Phys. Rev.} D {\bf 35} 3632

\bibitem{Motl03} Motl L and Neitzke A
2003 \textit{Adv. Theor. Math. Phys.} {\bf 7} 307

\bibitem{Andersson} Andersson N 1991 \textit{Class. Quantum Grav.} {\bf 10} L61

\bibitem{SHUK} Hod S and Keshet U 2006 \textit{Phys. Rev.} D {\bf 73} 024003


\end{thebibliography}
\end{document}